\documentclass[a4paper,twoside]{article}

\usepackage{epsfig}
\usepackage{subfigure}
\usepackage{calc}
\usepackage{amssymb}
\usepackage{amstext}
\usepackage{amsmath}
\usepackage{amsthm}
\usepackage{multicol}
\usepackage{pslatex}
\usepackage{apalike}

\usepackage{graphicx}

\usepackage{SCITEPRESS}     

\begin{document}

\title{Elephants, Donkeys, and Colonel Blotto\thanks{This project has received funding from the European Union's Horizon 2020 research and innovation programme under grant agreement No 732942.}}

\author{\authorname{Ivan P. Yamshchikov\sup{1}, Sharwin Rezagholi\sup{1}}
\affiliation{\sup{1}Max Planck Institute for Mathematics in the Sciences, Inselstrasse 22, 04103 Leipzig, Germany}
\email{\{ivan.yamshchikov, sharwin.rezagholi\}@mis.mpg.de}
}


\abstract{This paper employs a novel method for the empirical analysis of political discourse and develops a model that demonstrates dynamics comparable with the empirical data. Applying a set of binary text classifiers based on convolutional neural networks, we label statements in the political programs of the Democratic and the Republican Party in the United States\footnote{The donkey is a symbol of the Democratic Party and the elephant is a symbol of the Republicans.}. Extending the framework of the Colonel Blotto game by a stochastic activation structure, we show that, under a simple learning rule, the simulated game exhibits dynamics that resemble the empirical data.}

\onecolumn \maketitle \normalsize \vfill

\section{\uppercase{Introduction}}
\label{sec:introduction}

\noindent Electoral competition, the attempt of political actors to unite a large fraction of the voting population behind them, is of utmost importance in democratic systems. In this note we gather preliminary empirical insight into how political organizations divide their activities among political issues. We use data from political manifestos of English-speaking countries to classify the contents of US political manifestos into issues. We visualize the time-path of the fraction of statements dedicated to certain issues within the political programs. We use these empirical results to motivate a game-theoretic toy model of a bipartisan democracy, which we explore numerically. This model, under certain conditions, demonstrates behavior similar to the manifesto data. The numerical results suggest that this capability is due to our introduction of a stochastic activation structure of political issues in the voting population.

In the spirit of complex systems, we conceptualize political parties as adaptive agents in an uncertain environment, where they use a form of local optimization, while the outcome of their behavior depends on the opposing parties.

We first present our empirical results, then our toy model and its numerical analysis, and finally subsume and sketch possible future lines of research.

\section{\uppercase{Empirical data and motivation}}
\label{motivation}

\noindent It is not easy to measure the dynamics of political debates, although the Internet, with its wealth of data, has made it easier to measure some aspects of political discussions, see, for example, \cite{Jungherr}, \cite{Neuman} or \cite{Graham}. For a detailed discussion of the connection between political discussions, and the language used in them, we address the reader to the monographs of  \cite{Chilton} and \cite{Parker}.  For a discussion of opinion polarization we point to \cite{banischolbrich2017}. In this note we try to obtain an intuition on the dynamics of the political debate by looking at the programs of political parties. 

Programs of political parties are well-documented in developed countries. In this paper we focus on party programs from the United Kingdom, Canada, and the United States, which are collected in the Manifesto database \cite{manifesto}. We only work  with programs in the English language. The application of the proposed methods to other countries and languages is within the scope of future research. The Manifesto corpus consists of documents of different types. Some of them are annotated texts, where each (semi-) sentence is classified into one, and only one, category, others are plain texts, or scanned copies. There are seven categories in the Manifesto corpus:\footnote{We retain the categories of the dataset, but give them names that we hope are more precisely interpretable.}
\begin{itemize}
\item Foreign policy issues, with subcategories such as internationalism, foreign special relationships, or military issues. 
\item Freedom and law issues, with subcategories such as human rights, democracy, or constitutionalism.
\item Government issues, with subcategories such as centralization, administrative efficiency, or political corruption.
\item Economic policy issues, with subcategories such as technology, infrastructure, growth, or economic regulation.
\item Social policy issues, with subcategories such as equality, welfare state, or education.
\item Cultural policy issues, with subcategories such as national way of life, traditional morality, or multiculturalism.
\item Target groups issues, with subcategories such as appeal to labor groups, farmers, middle class, or minorities.
\end{itemize}
The categories listed above can be highly fragmented, but we believe that this level of generality is still suitable to analyze the macro-dynamics of political discourse.

The annotated data for the UK is available from 1997 to 2015, for the US the annotated data is available from 2004 to 2012, for Canada from 2011 to 2015. The time span for each country is extremely small, covers only a limited number of elections, and does not allow to see the long-term dynamics of political debate. The plain-text data from the US goes as far back as 1960. We are especially interested in the US, since we hypothesize that the dynamics of a bipartisan system are comparatively simple.
One needs to come up with a way to classify the non-annotated texts. One needs a classifier that associates each sentence in the non-annotated program to a category from the list above. Convolutional neural networks are a robust tool for tasks of this sort. Following the approach proposed by \cite{kim2014} we train seven binary classifiers on the annotated programs from the UK, the US, and Canada and apply these classifiers to the historical programs of the Democratic and the Republican Party of the United States. Details on the obtained classifiers are given in the Appendix.

Figure \ref{fig:discourse} shows the dynamics of seven categories in the political programs of Democrats and Republicans. Each subfigure shows the percentage of text in the program addressing a specific issue. These estimates are imperfect (the accuracy of each classifier is around 70\% on the test data), but since we apply the same classifiers to Republican and Democratic programs and Figure \ref{fig:discourse} depicts the percentage of sentences out of the total number of classified sentences, the visualization captures the qualitative dynamics of the discourse. Indeed Figure \ref{fig:discourse} provides several interesting insights.

\begin{figure*}[h!]
  \centering
   {\epsfig{file = 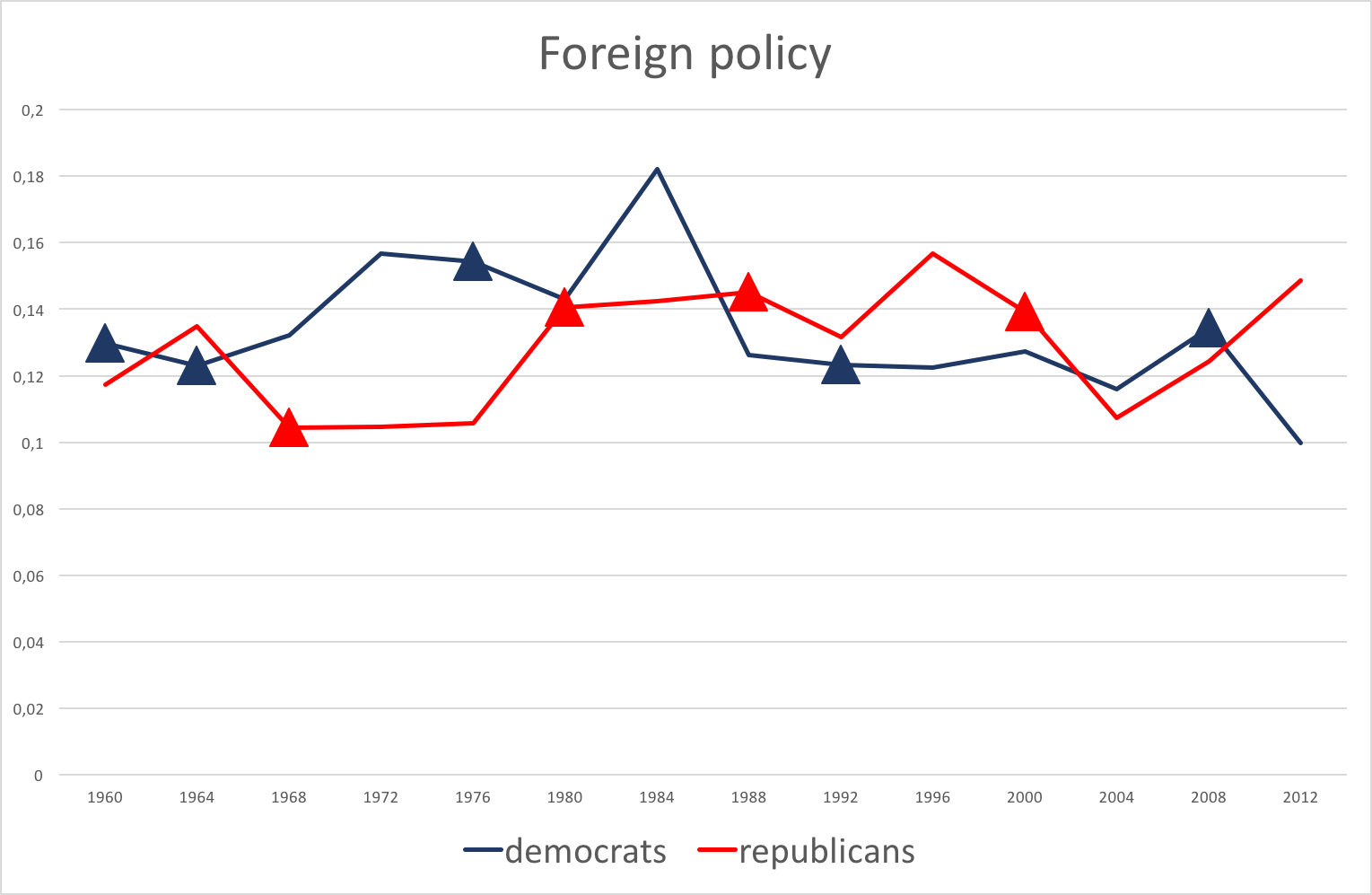, width = 7cm}}
   {\epsfig{file = 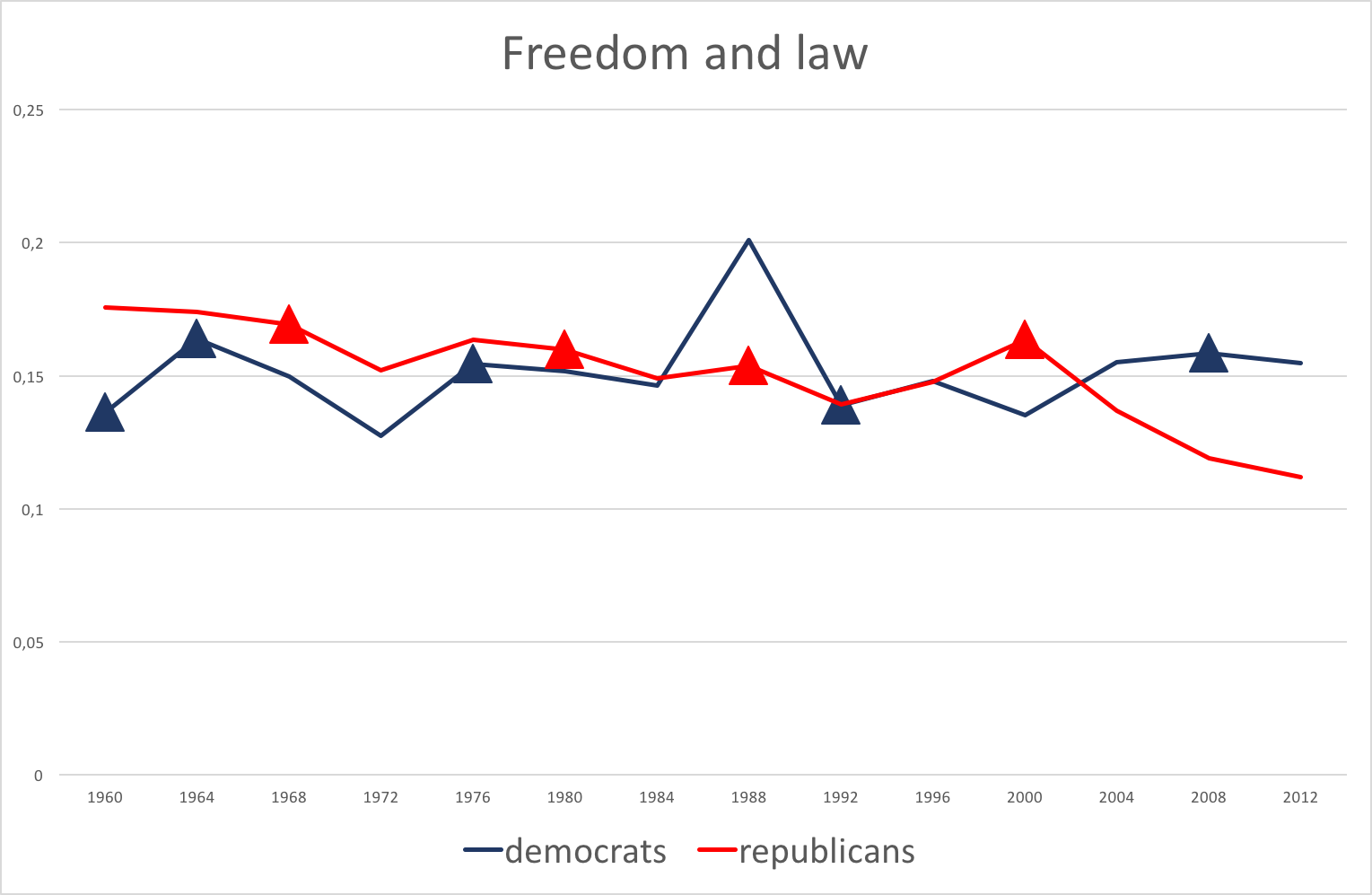, width = 7cm}}
   {\epsfig{file = 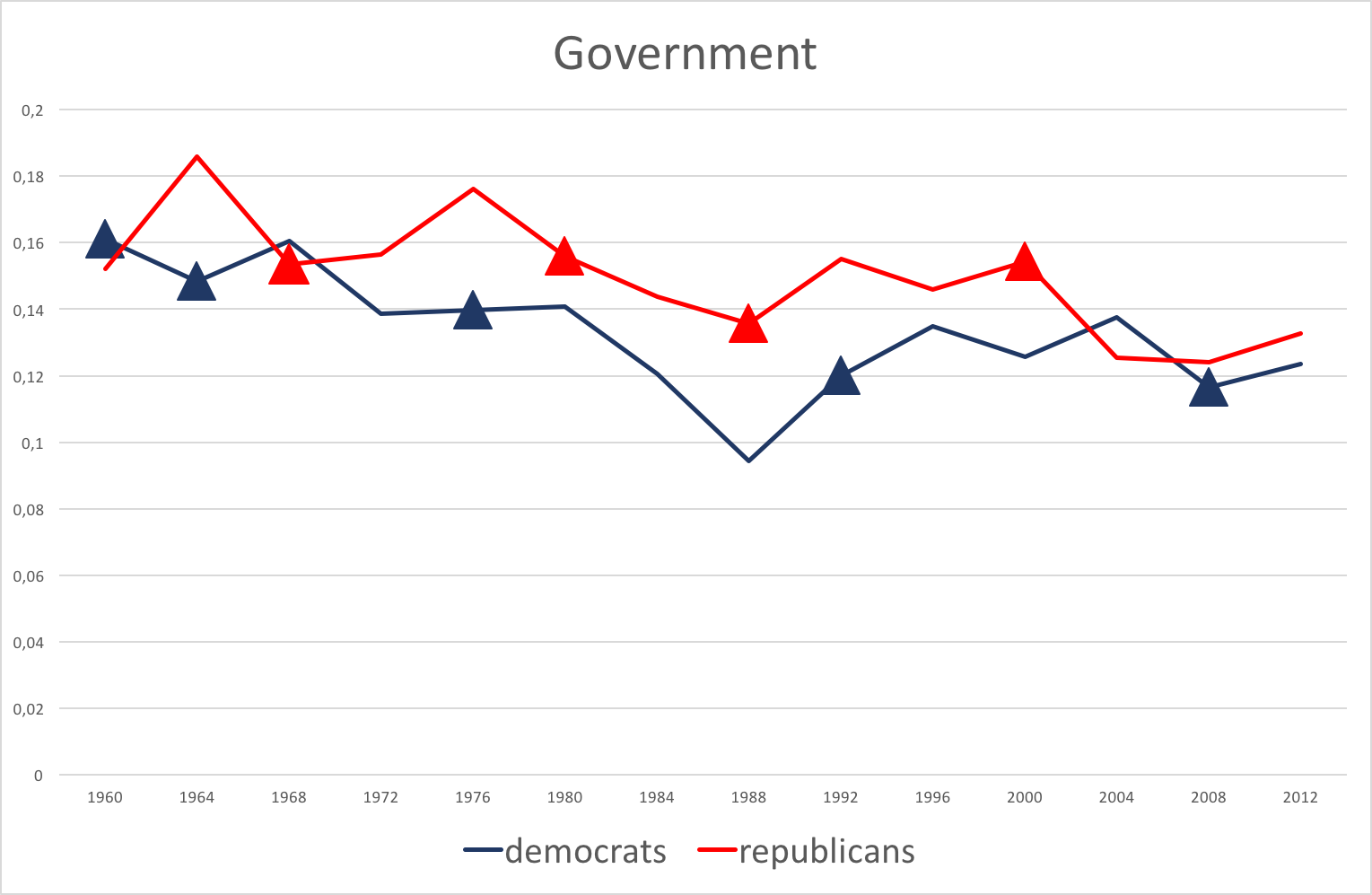, width = 7cm}}
   {\epsfig{file = 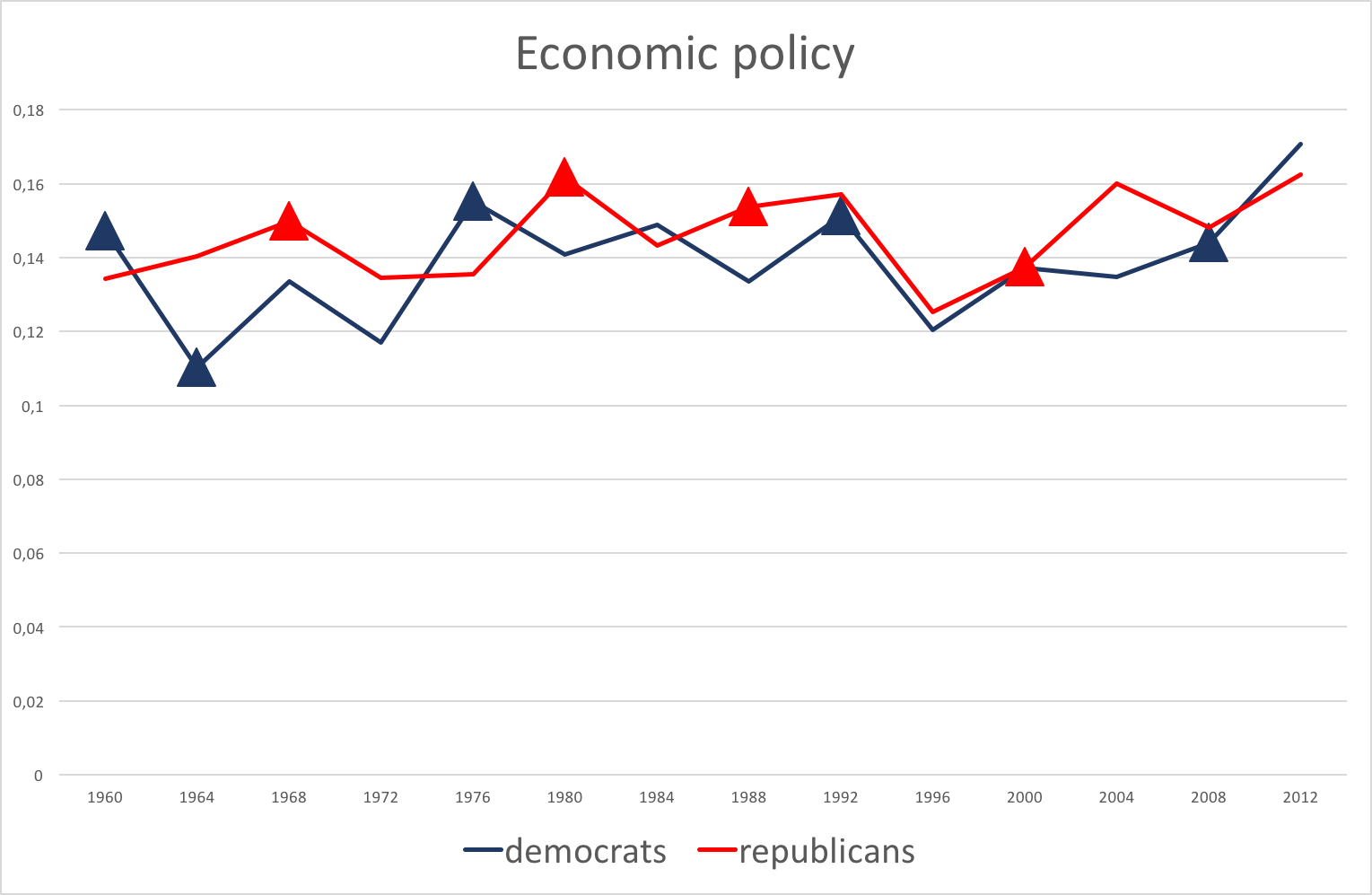, width = 7cm}}
   {\epsfig{file = 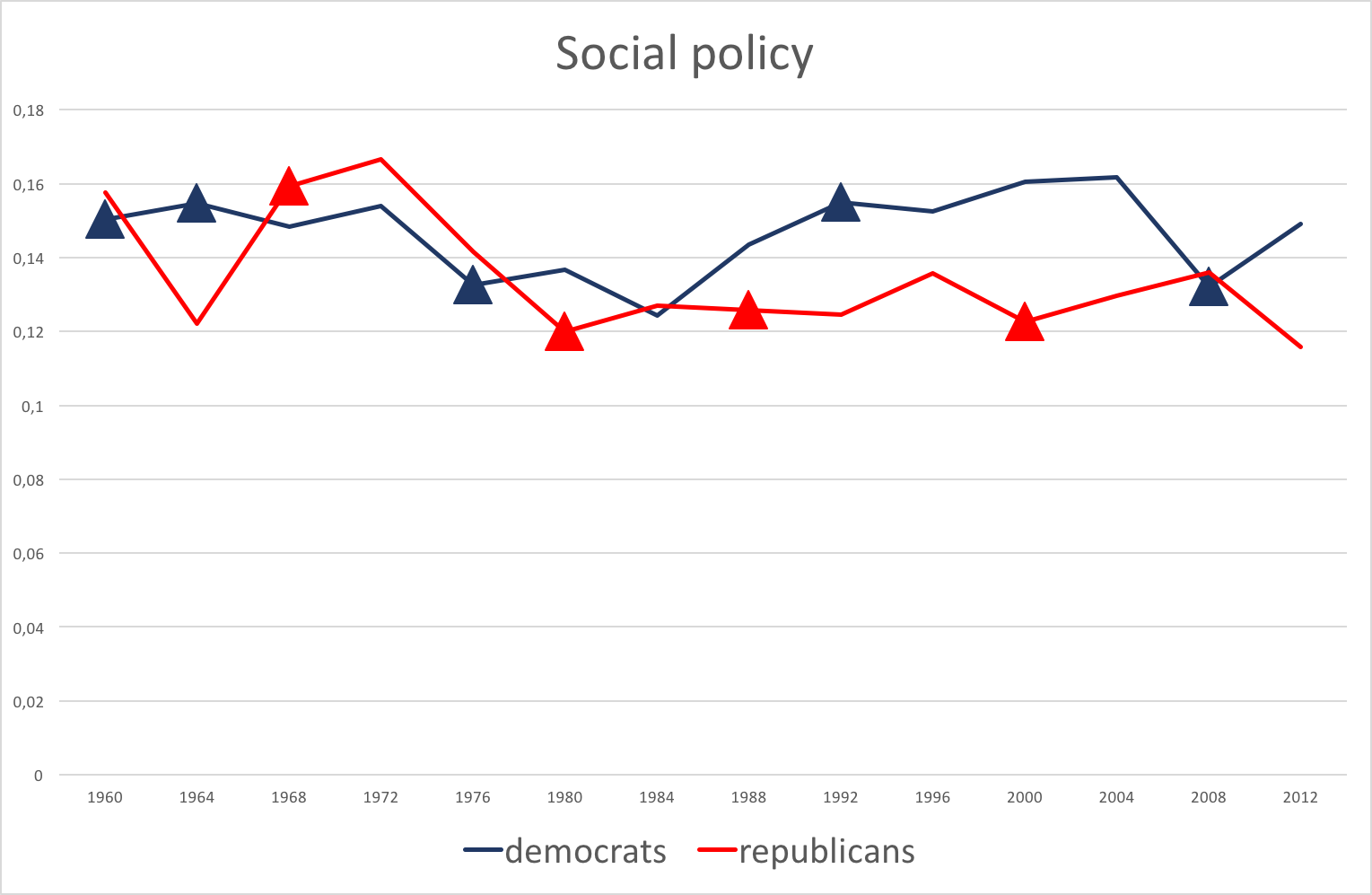, width = 7cm}}
   {\epsfig{file = 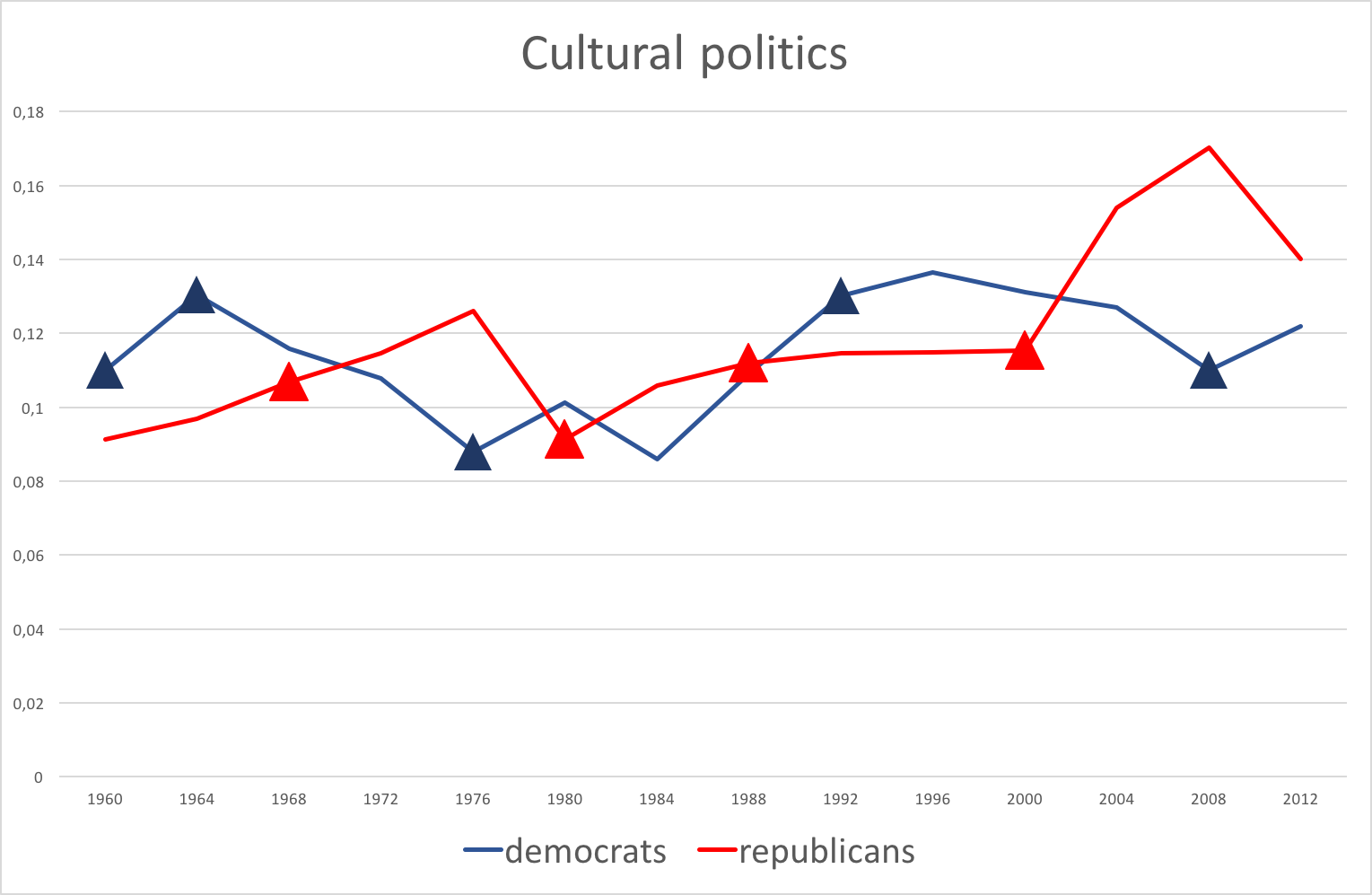, width = 7cm}}
   {\epsfig{file = 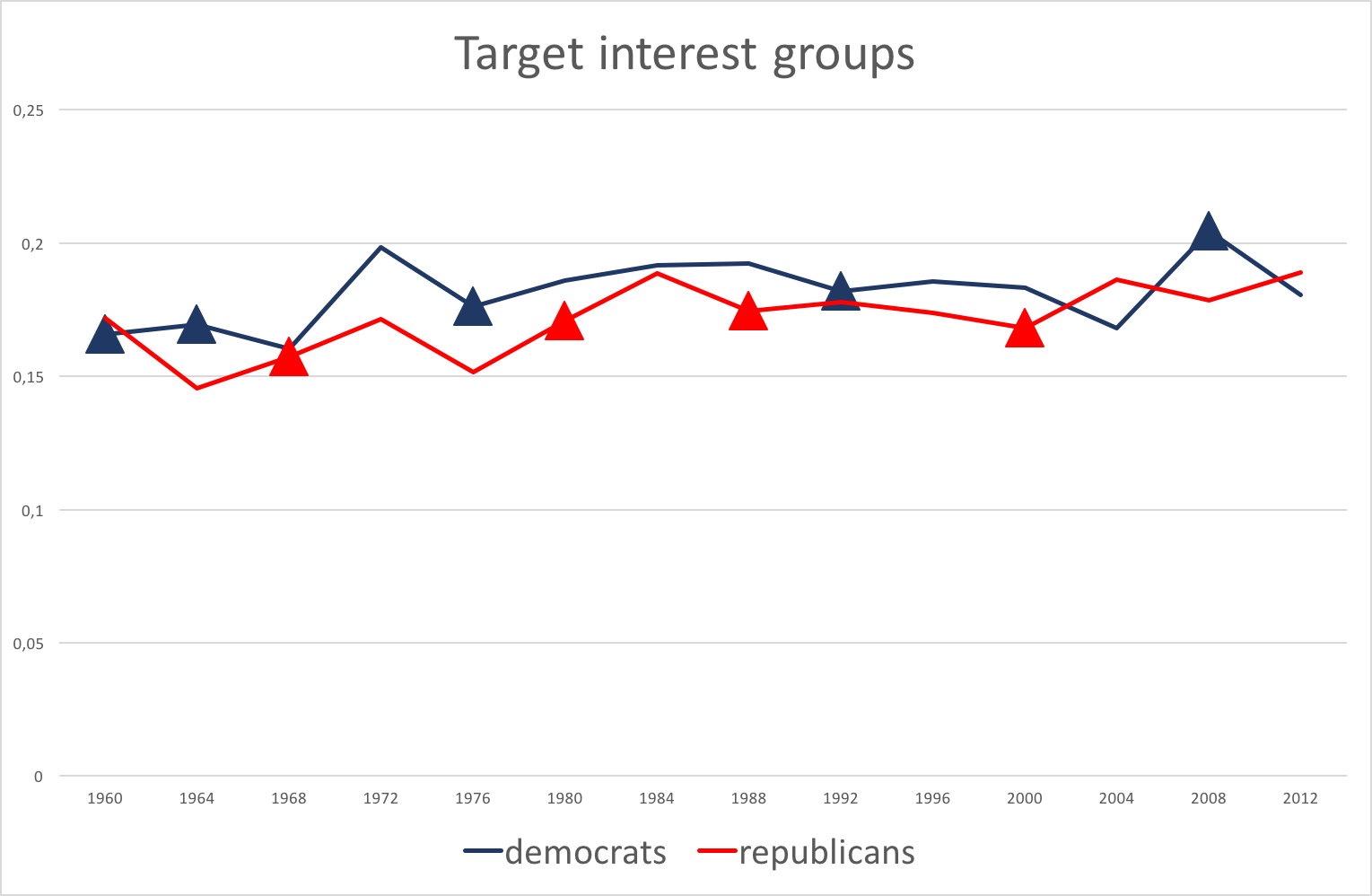, width = 7cm}}
   \vspace{5pt}
  \caption{Political discourse in the US, triangles mark wins of party candidates in presidential elections.}
  \label{fig:discourse}
 \end{figure*}

There are patterns which can be attributed to the actual history of the political process: In the eighties Democrats start to constantly pay more attention to social issues; In 2000 there is a rapid increase in Republican attention to culture-related issues. Other interesting patterns are surges of attention to government-related issues in the Republican programs of 1964 (after the death of Kennedy) and 1976 (after Watergate). It is interesting that the parties stay relatively close to each other on each issue. This is surprising, since we are talking about the percentage of the program devoted to a certain topic. If we look closely at presidential election results, we see another interesting pattern. If we give one point to the presidential candidate whose party has devoted a larger fraction of their program to a given issue, and give zero points to the candidate whose party devoted a smaller percentage to the issue, we see that, in the majority of the cases, the scores would be 4:3 or 3:4. In most cases the candidate whose score exceeds the opponent's wins the election. The scores are not always close to each other: Richard Nixon and  Bill Clinton win their first elections with 5:2 in 1972 and 1992 respectively. Ronald Reagan is reelected in 1984 with 5:2, on the verge of 6:1. 

These empirical results suggest the following explanation. As long as political parties compete across a number of different issues, the amount of resources that each party allocates to a given issue is proportional to its success among the voters that pay attention to the issue. The success of one party in a category can be modeled as 'the winner takes it all': The party that allocated more resources to the issue 'wins' it, while the opposing party 'looses' the issue. The party that 'gets the most issues' wins the election. Models with this structure are known in political science and economics as Colonel Blotto games. In the following sections we propose the extension of a Blotto game to model the dynamics of political discourse, run a simulation of this model, and briefly discuss its qualitative behavior.

\section{\uppercase{The toy model}}\label{sec:model}

\noindent Mathematical social scientists have used a game-theoretical model, originally conceived by \cite{borel1921} and typically referred to as the Colonel Blotto game, to model electoral competition. In the original setup of the game the military commander Blotto is tasked to divide his troops among a finite number of battlefields while his opponent does the same with his troops. A battlefield is won by the commander whose troops are in the majority on the respective battlefield. The commander who wins most of the battlefields wins the game.

While the Blotto game has been of interest to combinatorially inclined game theorists, it also became popular as a model of electoral competition \cite{myerson1993}. A thorough mathematical treatment of the game and a characterization of its equilibria is given by \cite{roberson2006}. For an analysis of redistributional politics on the basis of a Blotto model see \cite{laslierpicard2002}, who also provide an extensive bibliography. The equilibria of generalized versions of the classic Blotto game are studied by \cite{kovenockroberson2015}. A one-shot model with uncertainty that is similar to our setup is considered by \cite{osorio2013}.

We study a continuous Blotto game \cite{grosswagner1950}, where two players, henceforth called political parties, are endowed with a continuously divisible resource that they allocate between a finite number of battlefields, henceforth called political issues. One interpretation of the resource is as the media budget of the respective parties, which can be used to disseminate information, or even propaganda, on certain issues. Another interpretation of the resource is as the effort that politicians, or candidates for office, spend on furthering certain issues. Henceforth we will refer to the resource as the budget. We further allow asymmetry in the budgets, a stochastic activation structure of political issues (coupled with a limited form of history dependence), and consider political parties that use a myopic adaption rule. To our knowledge a comparable model has not yet been considered.

Consider two political parties, indexed by $i \in \{1,2\}$. The parties are endowed with budgets, modeled as real intervals $[0,b^i]$, $b^i >0$. The parties allocate their budgets to a finite set of political issues $\{1,...,n\}$. The game proceeds in stages, indexed by $t \in \mathbb{N}_0$. At $t=0$ the game is initialized with some initial allocation of resources $b^i \sigma^i (0)$, where $\sigma^i (0) \in \Delta \left( \{ 1,...,n \} \right)$ and $\Delta(A)$ denotes the probability simplex over the finite set $A$. Party $1$ wins the election in round $t$ if
\begin{equation}
\sum_{j=1}^n w_j (t) \left( b^1 \sigma_j^1 (t) - b^2 \sigma_j^2 (t) \right) > 0 \text{.}
\end{equation}
The winning condition for party $2$ is symmetric. Should the equation hold as an equality the election is drawn. The above equation encodes that party $1$ wins the election in period $t$ if it has allocated its resources such that the weighted sum of allocation differences is in its favor. The weighting vector $\left( w_1 (t) ,..., w_n (t) \right)$ in equation (1) is determined according to
\begin{equation}\label{eq:weights}
w_j (t) = \begin{cases}
1 \hspace{2pt} \text{with} \hspace{2pt} p_j  \left( 1 - \frac{1}{m} \sum_{l=t-m}^{t-1} w_j (l) \right) \\
\\
0 \hspace{2pt} \text{with} \hspace{2pt} 1 - p_j  \left( 1 - \frac{1}{m} \sum_{l=t-m}^{t-1} w_j(l) \right)
\end{cases}
\end{equation}
and is initialized as 
\begin{equation}
w_j (0) = \begin{cases}
1 \hspace{3pt} \text{with probability} \hspace{3pt} p_j \\
0 \hspace{3pt} \text{with probability} \hspace{3pt} 1 - p_j
\end{cases}
\end{equation}
where the vector $(p_1, ..., p_n) \in (0,1)^n$ parametrizes independent Bernoulli distributions. A more realistic model would incorporate a correlation structure between the activation of issues. This extension is a path for further research. The weighting vector consists of entries from $\{0,1\}$ which are redrawn independently in every period with a history dependent adjustment parametrized by $m \in \mathbb{N}$. The Bernoulli probability of activation of an issue is adjusted downwards according to the length of its activation period. In particular, if an issue has been activated for $m$ periods its activation probability vanishes entirely. This is meant to introduce some change of the political landscape: An issue cannot be important forever. Even if an issue is important with high probability (the $p_j$ parameter is close to one), it will sometimes be deactivated. Therefore changes of the political landscape provide a chance to parties with a small budget: Although the party with the higher budget can heavily cater to high-probability issues, the party with smaller budget can invest in a 'special interest' that might be sporadically activated. Our way of setting up the game is basically the plurality version of the Blotto game \cite{laslierpicard2002} with the addition of the weighting term.

We equip the parties with adaptive rules for action adjustment that can be described by the following procedure for each party
\begin{itemize}\label{rule}
\item Calculate new policy 
	\begin{equation}
		\sigma^i_j (t+1) = \sigma^i_j (t) + s^i (\sigma^o_j (t) - \sigma^i_j (t)) w (t),
	\end{equation}
	where $\sigma^o_j (t) $ is the policy of the opponent on the previous step.
\item Check that the new policy is positive $\sigma^i (t+1)>0$. Shift policies uniformly so that each $\sigma^i_j (t+1)$ is non negative
\item Normalize $\sigma^i (t+1)$ so that it satisfies the budget constraint  $\sum_{j=1}^n \sigma_j^i (t+1) = 1$
\end{itemize}
The parameter $s^i \in (0,1)$ is the speed of adjustment. This is an imitation-heuristic. The actions are initialized randomly. The adjustment rule is myopic in the sense that parties do not explicitly reason about the reaction of the political opponent and enact local, instead of global, optimization. We believe that this admits a fairly natural interpretation. Political organizations comprise many decentralized individual actors and organizations, political action groups, lobbyists, special interest groups, long-term media relations, and so on. We suggest the interpretation of the adjustment rule as the outcome of mass action.

While an analytical treatment of this type of model is of interest, we will now present preliminary numerical results in the next section.

\section{\uppercase{Simulations and discussion}}

\noindent We perform a number of simulations according to the strategy described in Equation \ref{rule}. We demonstrate that, despite its simplicity, our approach shows interesting results. The results of the simulation can be related to the empirical data described in Section \ref{motivation}. The number of issues is set to equal 7, the number of issues in the empirical data in Section \ref{motivation}.

\begin{figure*}[!ht]
  \centering
   {\epsfig{file = 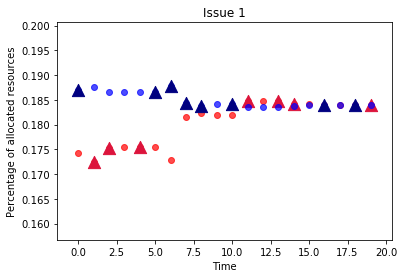, width = 7cm}}
   {\epsfig{file = 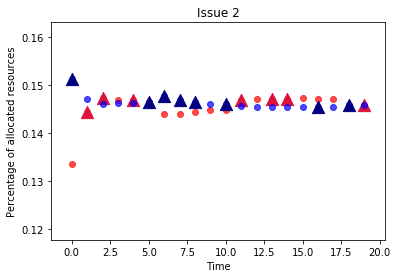, width = 7cm}}
   {\epsfig{file = 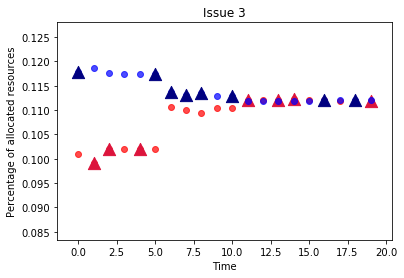, width = 7cm}}
   {\epsfig{file = 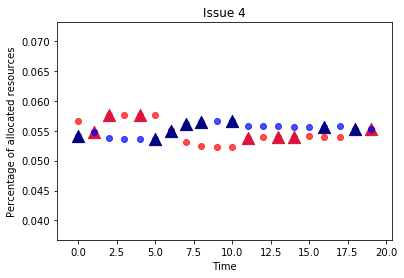, width = 7cm}}
   {\epsfig{file = 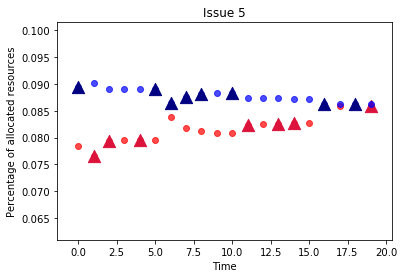, width = 7cm}}
   {\epsfig{file = 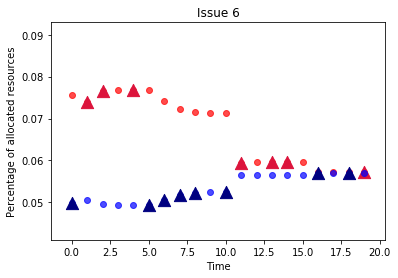, width = 7cm}}
   {\epsfig{file = 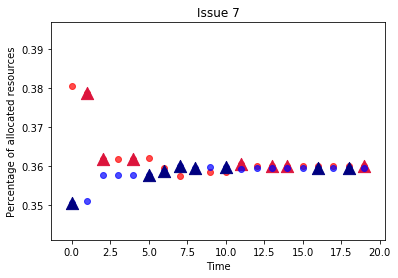, width = 7cm}}
  \caption{Simulation of political discourse across 7 issues and 20 elections, triangles mark wins of party candidates.}
  \label{fig:discoursesim}
   \vspace{-0.1cm}
 \end{figure*}

Before we discuss the results of our simulations, let us briefly elaborate on the parameters of the model. Since the importance of an issue in an election depends on a probability $p_j$, which can differ across issues, it makes sense to explore at least three different regimes of issue activation:
\begin{itemize}
\item All issues are activated fairly often,
\item All issues are activated rarely,
\item Some issues are activated much more often than the others.
\end{itemize}
The parameter $m$ in Equation \ref{eq:weights} can be interpreted as the 'inertia' of the voters' opinions: They still pay attention to issues that are in fact already obsolete. The speed of the adjustments $s^i$ controls how fast parties can reallocate their resources from an issue to another. The model allows every party to have a different budget $b^i$. These budgets are parameters corresponding to the collective efforts of the party to dominate in the political discussion. We assume the budgets to be equal in our simulations, however further exploration of the model without this constraint is of interest, especially in a multi-party setup. Figure \ref{fig:discoursesim} shows a simulation of 20 rounds of the game where some issues are activated much more often than the others. The triangles are denoting the winner of the election, as in the empirical results shown in Figure \ref{fig:discourse}. The path dependence is relatively small ($m=3$) and the speed of adjustment is relatively high ($s^1 = s^2 = 0.1$). One can see how the red player adjusts her stance on issue 6 on the 11th round and captures issues 1, 2 and 7.

\begin{figure}[!ht]
  \centering
   {\epsfig{file = 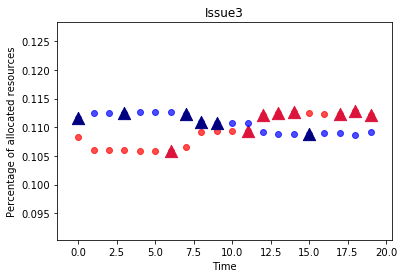, width = 7cm}}
  \caption{Red player captures and holds the third issue on the 12th round but it does not secure her position fully.}
  \label{fig:compdem}
   \vspace{-0.1cm}
 \end{figure}

As time passes the system approaches an equilibrium. Convergence is faster if parties can reallocate their resources quicker or if issues are activated more often. Rare feature-activation increases the fuzziness of the process. Generally, there are two distinct modes of the system: The first could be called competitive democracy, the second one-party-dominance with exogenous shocks. The competitive regime is characterized by political parties that are 'responsive' to the challenges (higher speeds of the players) and 'problems' that occur relatively rarely (issues get activated rarely, only a few are present at every given time). In this mode leadership moves from one party to the other often, as newly activated issues are captured by one of the parties. One can see the typical dynamics of one issue under these conditions in Figure \ref{fig:compdem}. The leadership changes several times and the winner keeps the position for longer periods of time. 

\begin{figure}[!ht]
  \centering
   {\epsfig{file = 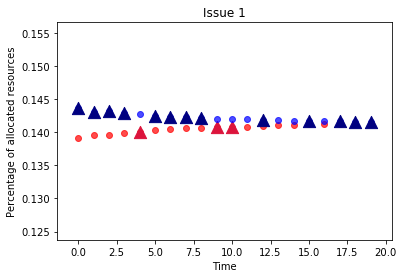, width = 7cm}}
  \caption{Blue player's domination fails due to random shocks rather than to issue capture.}
  \label{fig:auto}
   \vspace{-0.1cm}
 \end{figure}

The system enters the other regime if the problems are persistent (the $p_j$-values are relatively close to 1) and political parties are not very 'flexible' ($s^1$ and $s^2$ are small). In this mode one party captures the leadership due to a strong initial position and loses it only for short periods of time, due to the exogenous issue-deactivation, rather than due to the actions of the opponent. The parties are not flexible and can not aptly respond to the challenges that they face. The opposition can capture power for a short period of time due to exogenous shock but the 'usual' political order is soon restored.  In Figure \ref{fig:auto} one can see the dynamics of an issue in this regime. Since most of the issues are active most of the time and the speed $s^i$ of the parties is low, the parties have little chance to reallocate resources in a way that would allow them to win and keep winning for some time. Changes of leadership occur due to exogenous shocks and the power goes back to the dominant player (the blue one in Figure \ref{fig:auto})  after a short period of time.

\section{\uppercase{Conclusion}}

\noindent This note is a foray that could be taken further and investigated in greater detail. First, using data from the Manifesto project database \cite{manifesto}, we have trained 7 binary classifiers based on a pre-trained convolutional neural network built according to \cite{kim2014}. The obtained classifiers have demonstrated relatively high precision on the test data (precision was around 70\%) and allowed us to label historical political programs of Democrats and Republicans going as far back as 1960. Second, an analysis of the labeled historical data supports the hypothesis that the dynamics of political discourse follow a form of the Colonel Blotto game. This opinion was previously voiced in relation to different aspects of the political process (\cite{washburn}, \cite{roberson}, \cite{kovenockroberson2015}) but this is, to our knowledge,  the first paper where this hypothesis is complemented by empirical data. Third, a dynamic stochastic model of electoral competition with learning and history dependence is provided. It is an extension of the Colonel Blotto game that we also believe to be novel. The extension of the model to multiple parties, and the empirical justification of such an extension by data from a multi-party electoral system, are a matter for future research. Finally, a simulation with a simple policy update-rule is able to produce data that shares the qualitative properties of the empirical findings. Models of the type presented here may be developed further to provide insights into, at least the qualitative, phenomena in electoral competition. The ultimate aim of social science is of course prediction, but we do not believe that this is a realistic target. Still, providing tractable models of electoral competition might allow us to develop deeper insight into possible futures and to identify the key mechanisms of political change.

\section*{\uppercase{Acknowledgements}}
\noindent The authors are grateful to Eckehard Olbrich and Sven Banisch for fruitful and interesting discussions and their advice and support.

\vfill
\bibliographystyle{apalike}
{\small
\bibliography{EDCB}{}}

\section*{\uppercase{Appendix}}

\noindent In \cite{kim2014} it is shown that, despite little tuning of hyperparameters, a simple neural network with one layer of convolution performs remarkably well on sentence-classification tasks, when provided with unsupervised pre-training of word vectors. We followed the proposed method and built seven binary classifiers where we used labeled sentences from \cite{manifesto} as positive examples and randomly selected sentences with different label from the same data set as negative ones. The accuracy results of the vanilla classifiers on test data are provided in Table \ref{tab:example1}.

\vspace{5pt}
\begin{table}[h]
\caption{Obtained binary classifiers.}\label{tab:example1} \centering
\begin{tabular}{ccc}
  Category & Size of train & Test accuracy \\
  \hline
  Foreign policy & 8486 & 69.3 \\
  Freedom and law & 4442 & 70.1 \\
  Government & 9554 & 70.5 \\
  Economic policy  & 22013 & 71.4 \\
  Social policy & 26340 & 72.1 \\
  Cultural policy & 9136 & 69.8 \\
  Target groups & 9256 & 69.3 \\
  \hline
\end{tabular}
\end{table}

\vfill
\end{document}